\documentclass[aps,prl,preprint,tightenlines,superscriptaddress,showpacs,byrevtex]{revtex4}
\usepackage{graphicx} 
\usepackage{dcolumn}  
\usepackage{epsfig}

\graphicspath{{ps}}

\newcommand{\BB}{\ensuremath{B\overline{B}}}
\newcommand{\Bbar}{\ensuremath{\overline{B}}}
\newcommand{\Bbaro}{\ensuremath{\overline{B}{}^0}}

\newcommand{\de}{\ensuremath{\Delta E}}
\newcommand{\mb}{\ensuremath{M_{\mbox{\scriptsize bc}}}}
\newcommand{\etap}{\ensuremath{\eta^\prime}}
 
\newcommand{\BF}{\ensuremath{{\mathcal B}}}

\newcommand{\LK}{\ensuremath{{\cal L}}}
\newcommand{\RK}{\ensuremath{{\cal R}_K}}
\newcommand{\FD}{\ensuremath{{\cal F}}}
\newcommand{\LR}{\ensuremath{{\cal R_L}}}
\newcommand{\hel}{\ensuremath{{\cal H}}}
\newcommand{\cost}{\ensuremath{\cos\theta_T}}
\newcommand{\cosb}{\ensuremath{\cos\theta_B}}
\newcommand{\sperp}{\ensuremath{S_\perp}}
\newcommand{\epp}{\ensuremath{\eta' \to \eta \pi^+ \pi^-}}
\newcommand{\erg}{\ensuremath{\eta' \to \rho^0 \gamma}}
\newcommand{\ebeam}{\ensuremath{E_{\mbox{\scriptsize beam}}}}

\newcommand{\cs}{\ensuremath{/c^2}}
\newcommand{\beq}{\begin{eqnarray}}
\newcommand{\eeq}{\end{eqnarray}}
\newcommand{\eff}{\ensuremath{\epsilon}}
 
\newcommand{\ACP}{\ensuremath{A_{CP}}}

\newcommand{\btepk}{\ensuremath{B \to \etap K}}
\newcommand{\bteppi}{\ensuremath{B \to \etap \pi}}
\newcommand{\btepkp}{\ensuremath{B^+ \to \etap K^+}}
\newcommand{\bteppip}{\ensuremath{B^+ \to \etap \pi^+}}
\newcommand{\btepks}{\ensuremath{B^0 \to \etap K_S^0}}

\begin{document}


\preprint{\vbox{ \hbox{   }
                 \hbox{BELLE-CONF-0563}
                 \hbox{LP2005-198}
                 \hbox{EPS05-538} 
                 \hbox{hep-ex nnnn}
}}

\title{ \quad\\[0.5cm]  Study of $B \to \etap h$}


\affiliation{Aomori University, Aomori}
\affiliation{Budker Institute of Nuclear Physics, Novosibirsk}
\affiliation{Chiba University, Chiba}
\affiliation{Chonnam National University, Kwangju}
\affiliation{University of Cincinnati, Cincinnati, Ohio 45221}
\affiliation{University of Frankfurt, Frankfurt}
\affiliation{Gyeongsang National University, Chinju}
\affiliation{University of Hawaii, Honolulu, Hawaii 96822}
\affiliation{High Energy Accelerator Research Organization (KEK), Tsukuba}
\affiliation{Hiroshima Institute of Technology, Hiroshima}
\affiliation{Institute of High Energy Physics, Chinese Academy of Sciences, Beijing}
\affiliation{Institute of High Energy Physics, Vienna}
\affiliation{Institute for Theoretical and Experimental Physics, Moscow}
\affiliation{J. Stefan Institute, Ljubljana}
\affiliation{Kanagawa University, Yokohama}
\affiliation{Korea University, Seoul}
\affiliation{Kyoto University, Kyoto}
\affiliation{Kyungpook National University, Taegu}
\affiliation{Swiss Federal Institute of Technology of Lausanne, EPFL, Lausanne}
\affiliation{University of Ljubljana, Ljubljana}
\affiliation{University of Maribor, Maribor}
\affiliation{University of Melbourne, Victoria}
\affiliation{Nagoya University, Nagoya}
\affiliation{Nara Women's University, Nara}
\affiliation{National Central University, Chung-li}
\affiliation{National Kaohsiung Normal University, Kaohsiung}
\affiliation{National United University, Miao Li}
\affiliation{Department of Physics, National Taiwan University, Taipei}
\affiliation{H. Niewodniczanski Institute of Nuclear Physics, Krakow}
\affiliation{Nippon Dental University, Niigata}
\affiliation{Niigata University, Niigata}
\affiliation{Nova Gorica Polytechnic, Nova Gorica}
\affiliation{Osaka City University, Osaka}
\affiliation{Osaka University, Osaka}
\affiliation{Panjab University, Chandigarh}
\affiliation{Peking University, Beijing}
\affiliation{Princeton University, Princeton, New Jersey 08544}
\affiliation{RIKEN BNL Research Center, Upton, New York 11973}
\affiliation{Saga University, Saga}
\affiliation{University of Science and Technology of China, Hefei}
\affiliation{Seoul National University, Seoul}
\affiliation{Shinshu University, Nagano}
\affiliation{Sungkyunkwan University, Suwon}
\affiliation{University of Sydney, Sydney NSW}
\affiliation{Tata Institute of Fundamental Research, Bombay}
\affiliation{Toho University, Funabashi}
\affiliation{Tohoku Gakuin University, Tagajo}
\affiliation{Tohoku University, Sendai}
\affiliation{Department of Physics, University of Tokyo, Tokyo}
\affiliation{Tokyo Institute of Technology, Tokyo}
\affiliation{Tokyo Metropolitan University, Tokyo}
\affiliation{Tokyo University of Agriculture and Technology, Tokyo}
\affiliation{Toyama National College of Maritime Technology, Toyama}
\affiliation{University of Tsukuba, Tsukuba}
\affiliation{Utkal University, Bhubaneswer}
\affiliation{Virginia Polytechnic Institute and State University, Blacksburg, Virginia 24061}
\affiliation{Yonsei University, Seoul}
  \author{K.~Abe}\affiliation{High Energy Accelerator Research Organization (KEK), Tsukuba} 
  \author{K.~Abe}\affiliation{Tohoku Gakuin University, Tagajo} 
  \author{I.~Adachi}\affiliation{High Energy Accelerator Research Organization (KEK), Tsukuba} 
  \author{H.~Aihara}\affiliation{Department of Physics, University of Tokyo, Tokyo} 
  \author{K.~Aoki}\affiliation{Nagoya University, Nagoya} 
  \author{K.~Arinstein}\affiliation{Budker Institute of Nuclear Physics, Novosibirsk} 
  \author{Y.~Asano}\affiliation{University of Tsukuba, Tsukuba} 
  \author{T.~Aso}\affiliation{Toyama National College of Maritime Technology, Toyama} 
  \author{V.~Aulchenko}\affiliation{Budker Institute of Nuclear Physics, Novosibirsk} 
  \author{T.~Aushev}\affiliation{Institute for Theoretical and Experimental Physics, Moscow} 
% ...
\collaboration{The Belle Collaboration}
\noaffiliation

\begin{abstract}
We report improved measurements of exclusive two-body charmless hadronic 
$B$ meson decays
$B \to \etap h$, where $h$ is a charged kaon or pion or a $K^0$. 
These results are obtained from a data sample that contains 386
million $B\bar{B}$ pairs collected at the $\Upsilon(4S)$ resonance,
with the Belle detector at the KEKB asymmetric energy $e^+ e^-$
collider.
We measure
${\BF}(B^0\to \etap K^0) = (56.6 \, ^{+3.6}_{-3.5} \, \pm 3.3)
\times 10^{-6}$,
${\BF}(B^+\to \etap K^+) = (68.6 \, \pm 2.1 \, \pm 3.6)\times 10^{-6}$
and
${\BF}(B^+\to \etap \pi^+) = (1.73 \, ^{+0.69}_{-0.63} \,
\pm 0.12)\times 10^{-6}$, 
where the first and second errors are statistic and systematic, respectively.
The $CP$ asymmetries in the charged modes are measured and no evidence for 
direct $CP$ violation is found. We measure 
$\ACP (B^{\pm}\to \etap K^{\pm}) = 0.03 \pm 0.03 \pm 0.02$ and 
$\ACP (B^{\pm}\to \etap \pi^{\pm}) = 0.15 ^{+0.39}_{-0.38} \, ^{+0.02}_{-0.06}$.
\end{abstract}


\maketitle

\tighten

{\renewcommand{\thefootnote}{\fnsymbol{footnote}}}
\setcounter{footnote}{0}

\begin{figure}[hbtp]
\begin{center}
\unitlength1.0cm
\includegraphics[clip,angle=0,width=2.8in]{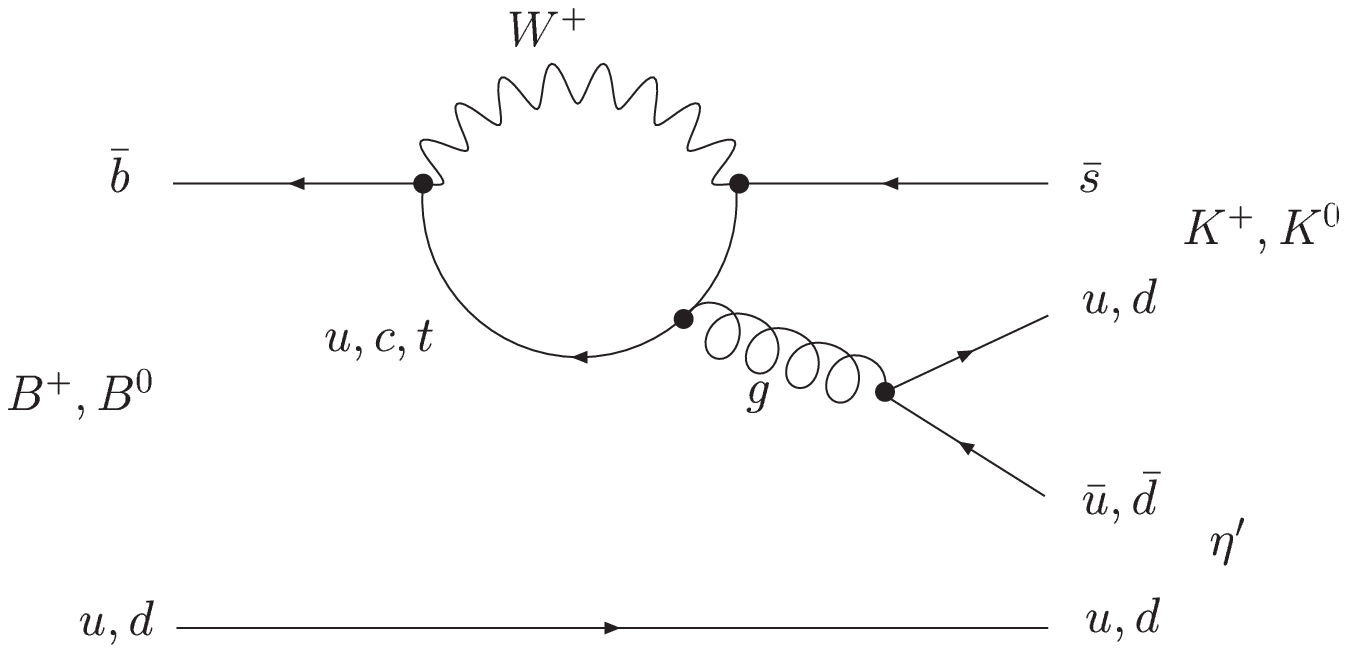}
\includegraphics[clip,angle=0,width=3.2in]{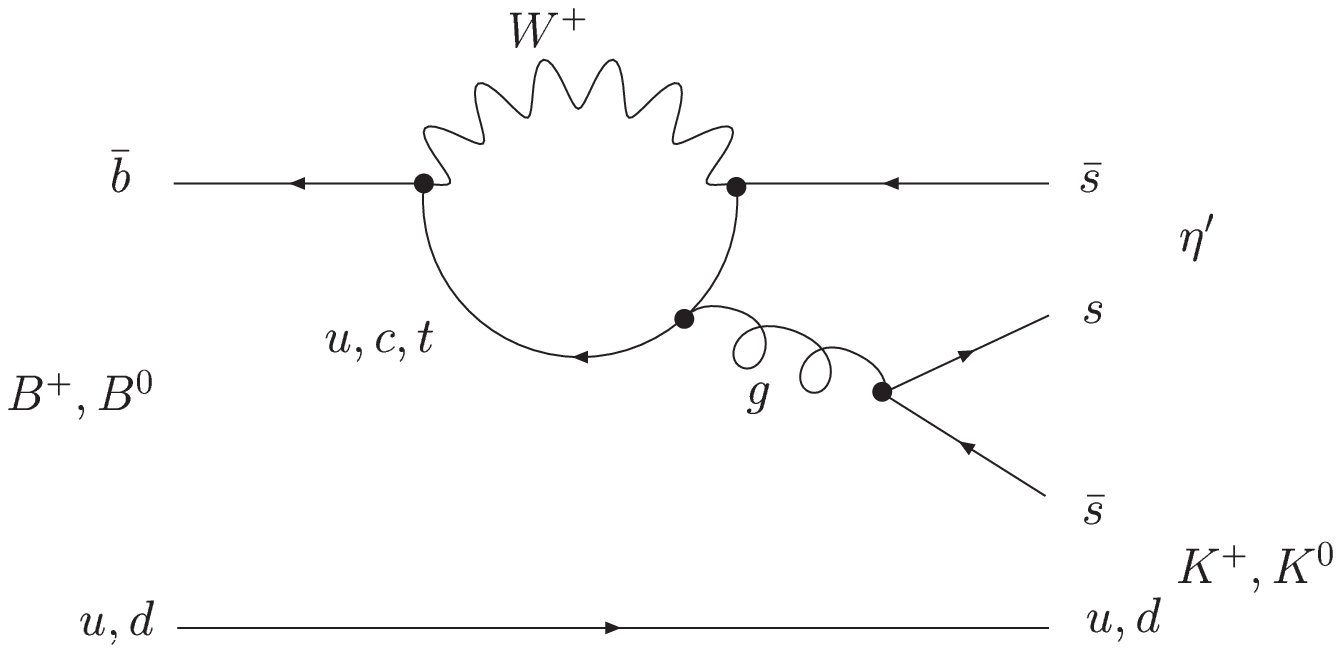}
\includegraphics[clip,angle=0,width=2.5in]{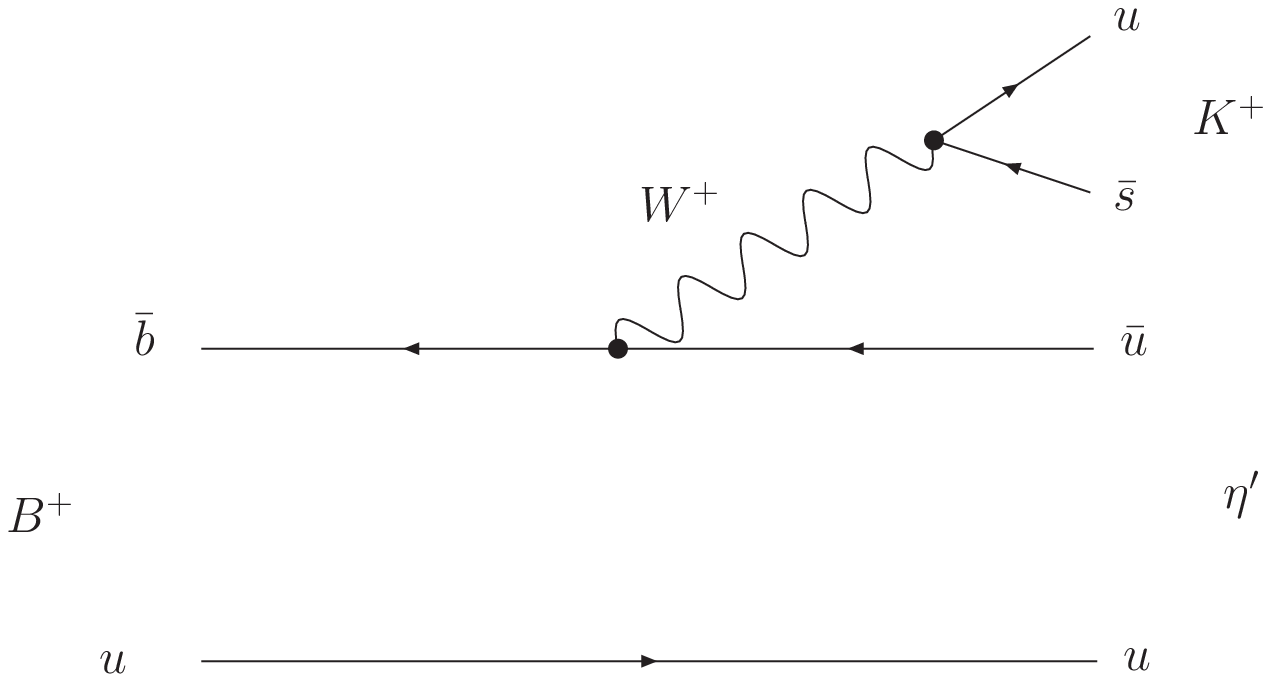}
\includegraphics[clip,angle=0,width=3.0in]{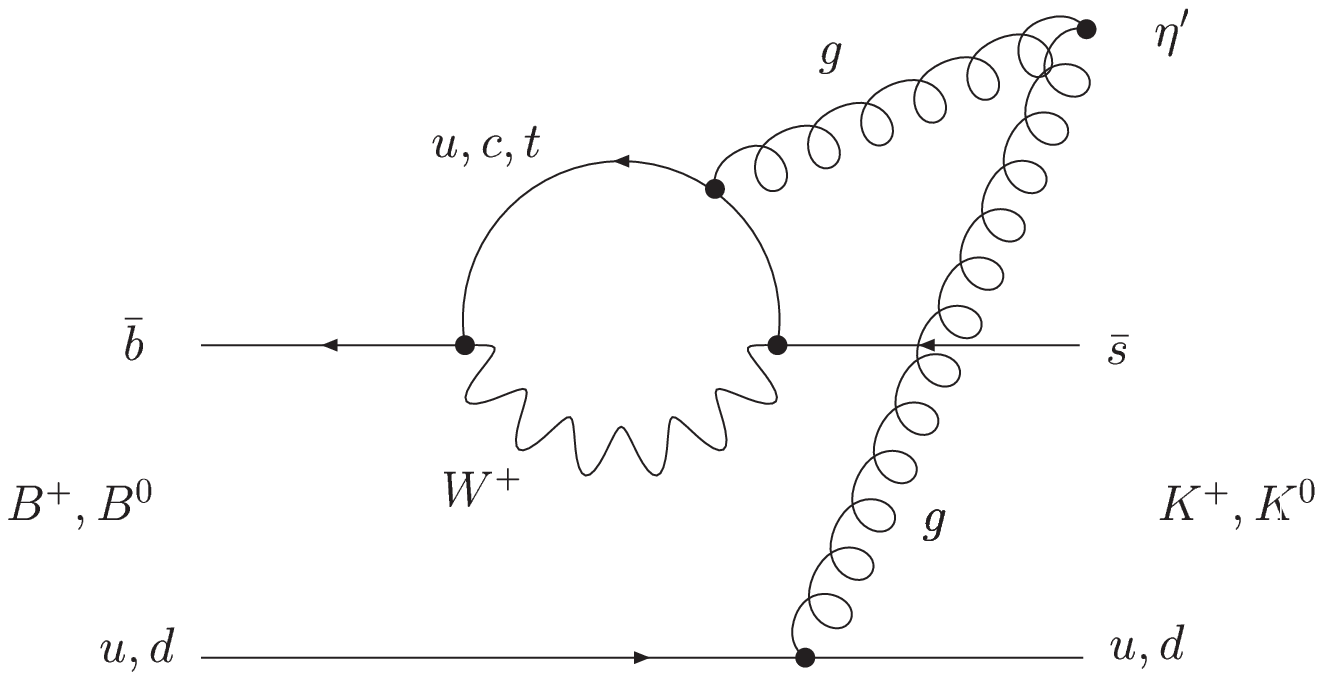}
\put(-14.0,4.8){\small{\sf\shortstack[c]{(a)}}}
\put(-6.9,4.8){\small{\sf\shortstack[c]{(b)}}}
\put(-14.0,-0.6){\small{\sf\shortstack[c]{(c)}}}
\put(-6.9,-0.6){\small{\sf\shortstack[c]{(d)}}}
\medskip
\medskip
\medskip
\caption{\label{fig:Feyn} Feynman diagrams describing the decays
\btepk$^{+/0}$: (a), (b) internal penguins, (c) external tree, (d)
flavour-singlet penguin.}
\end{center}
\end{figure}

The properties of the decay $B\to \etap X_s$ are still to be understood by
theory.
The channel \btepk{} has the largest branching
fraction of all charmless hadronic B decay modes. In the Standard
Model (SM) the decay \btepk{} is thought to proceed dominantly via
gluonic penguin processes~\cite{Kou:1999tt,Atwood:1997bn,Grossman:1996ke}. 
The diagrams~\cite{CC} that contribute are shown in Fig.~\ref{fig:Feyn}. 
The measured branching fractions for
\btepk{}~\cite{Richichi:1999kj,Abe:2001pf,Aubert:2003bq} are
larger than expectations from the generalized factorization
approach~\cite{Ali:1998eb,Chen:1999nx,Kou:2001pm,Chiang:2001ir}.
This has led to speculations that SU(3)-singlet couplings
unique to the \etap{} meson or new
physics~\cite{Xiao:2001uh,Dutta:2002as,Khalil:2003bi,Dariescu:2004bs} 
contribute to the amplitude.

The \bteppip{} decay proceeds via similar processes as \btepk{} with $b\to u
(d)$ tree or penguin decays. 
However, for the latter all contributions are additive, while for
the former tree and penguin contributions interfere destructively.

In the SM, 
direct $CP$ violation arises from the interference of two
or more amplitudes with different strong and weak phases~\cite{Kramer:1993yu}.
Many charmless hadronic $B$ meson decays contain both
tree and penguin amplitudes and provide a rich sample for 
direct $CP$ violation studies.
The decay \bteppi{} is expected to have comparable contributions
from both tree and penguin diagrams, and therefore may have a large
asymmetry.
For the decay processes dominated by one single amplitude in the SM,
a non-zero $CP$ asymmetry may indicate additional amplitudes in the decay
and hence provide a hint of new physics.
This is also the case for \btepk{}, which is thought to 
proceed entirely through the penguin amplitude 
and no $CP$ asymmetry is expected.

With the convention that a $\Bbar$ meson contains a $b$ quark,
the direct $CP$ asymmetry in a $B\to f$ decay is defined as 
$$\ACP = \frac{\Gamma(\Bbar\to\bar{f}) - \Gamma(B\to f)}
          {\Gamma(\Bbar\to\bar{f}) + \Gamma(B\to f) },
$$
where $f$ is the final state and $\bar f$ its $CP$ conjugate.
In the experiment we measure the branching fraction $\BF(B\to f)$
and $\BF(\Bbar\to \bar f)$, which are proportional to the 
partial widths.

In this paper, we update the measurements of 
$B\to\etap K^+,\etap K^0,\etap \pi^+$ with a sample 35 times larger than our
previous dataset~\cite{Abe:2001pf}. 
We also report 
the charge asymmetry of self-tagged decays.
All results are based on a data sample that
contains 386 million $B\overline{B}$ pairs 
collected  with the Belle detector at the KEKB asymmetric-energy
$e^+e^-$ (3.5 on 8~GeV) collider~\cite{KEKB}.
KEKB operates at the $\Upsilon(4S)$ resonance 
($\sqrt{s}=10.58$~GeV) with a peak luminosity that exceeds
$1.5\times 10^{34}~{\rm cm}^{-2}{\rm s}^{-1}$.

The Belle detector is a large-solid-angle magnetic
spectrometer that
consists of a silicon vertex detector (SVD),
a 50-layer central drift chamber (CDC), an array of
aerogel threshold \v{C}erenkov counters (ACC), 
a barrel-like arrangement of time-of-flight
scintillation counters (TOF), and an electromagnetic calorimeter (ECL)
comprised of CsI(Tl) crystals located inside 
a super-conducting solenoid coil that provides a 1.5~T
magnetic field.  An iron flux-return located outside of
the coil is instrumented to detect $K_L^0$ mesons and to identify
muons (KLM).  The detector
is described in detail elsewhere~\cite{Belle}.
Two inner detector configurations were used. A 2.0 cm beampipe
and a 3-layer silicon vertex detector was used for the first sample
of 152 million $B\bar{B}$ pairs (Set I), while a 1.5 cm beampipe, a 4-layer
silicon detector and a small-cell inner drift chamber were used to record  
the remaining 234 million $B\bar{B}$ pairs (Set II)~\cite{Ushiroda}.

%
The \etap{} 
mesons are reconstructed via either \epp, with $\eta \to
\gamma \gamma$, or \erg . $K^0$ mesons are reconstructed only
 via $K_S^0 \to \pi^+\pi^-$.

Primary
charged tracks were selected with $dr < 0.5$ cm and $|dz| < 2.0$ cm,
where $dr$ and $dz$ are the impact parameters perpendicular to and along the
beam axis, with respect to the run dependent interaction point (IP). 

Likelihoods for kaon and pion hypotheses, $\LK_K$ and $\LK_{\pi}$,
are obtained by combining information from the CDC ($dE/dx$), ACC and TOF 
systems. The likelihood ratio $\RK = \LK_{K}/(\LK_{\pi}+\LK_{K})$ ranges 
between $0$ (pion-like) and $1$ (kaon-like) and we require 
$\RK > 0.6$ for kaons and $\RK <0.4$ for
pions from $K_S^0$,
which keeps about
86\% of the kaon/pion candidates,
and $\RK <0.9$ for pion candidates from the \etap{} decay with an efficiency of
about 98\%. 
Prompt pions in \bteppip{}
are required to pass the tight selection, $\RK < 0.1$, which has an efficiency 
of about 85\% and a kaon fake rate of about 4\%.


The $\eta$ meson is reconstructed from two photons, each with an energy of at
least $50$ MeV. The $\eta$ mass window is chosen to be 
$0.5$ GeV\cs$ < M(\gamma\gamma) < 0.57$ GeV\cs, which
corresponds to $+2/-3$ standard deviations ($\sigma$) from the
nominal $\eta$ mass given by Particle Data Group (PDG)~\cite{bib:PDG04}.

We retain $\rho^0$ candidates in the mass range $550$ MeV\cs$ < M(\pi^+\pi^-) <
870$ MeV\cs, where $M(\pi^+\pi^-)$ is the $\rho^0$ candidate mass. 
We require the transverse momenta of the daughter pions to satisfy
the requirement, $P_T(\pi) > 0.2$ GeV$/c$. This suppresses around $40 \% 
$ of background while retaining $86 \%
$ of the signal. 

The requirements for the \etap{} candidate mass depend on the
decay channel. 
For \btepk{} decays, we select $\etap$ candidates within
$\pm 3.4\sigma$ window for $\epp$
(0.945 GeV/$c^2 < M(\eta\pi\pi) < $ 0.97 GeV/$c^2$) and
$\pm 3\sigma$ for $\etap \to \rho^0 \gamma$
(0.935 GeV/$c^2 < M(\rho\gamma) < $ 0.975 GeV/$c^2$).
For $\bteppip$ decay, the $\etap$
mass windows are tightened to $\pm 2.5\sigma$ (0.95 GeV/$c^2
< M(\eta\pi\pi) < $ 0.965 GeV/$c^2$ and 0.941 GeV/$c^2 <
M(\rho\gamma) < $ 0.97 GeV/$c^2$).
A weak requirement on the $\eta$ meson decay angular distribution, 
$h(\eta) < 0.97$,
has been applied for the \epp{} mode to suppress combinatorial background, 
where $h(\eta)$ is the cosine of the angle between the $\etap$
momentum and the direction of one of the decay photons
in the $\eta$ rest frame.

$K_S^0$ candidates are reconstructed from a pair of oppositely
charged particles with invariant mass within the range
$485$ MeV\cs $< M(\pi^+\pi^-) < 510$ MeV\cs.
We require the vertex of a $K_S^0$ to be well reconstructed and 
displaced from the interaction point
and the $K_S^0$ momentum direction to be consistent with the $K_S^0$ 
flight direction. 

$B$ meson candidates are then reconstructed combining an $\etap$ meson and 
one of the $h$ candidates. Two kinematic variables are used to extract the 
$B$ meson
signal: the energy difference $\de = E_B-\ebeam$ and the beam-energy constrained
mass $\mb =\sqrt{\ebeam^2 - P_B^2}$, where \ebeam{} is the beam energy and $E_B$
and $P_B$ are the reconstructed energy and momentum of the $B$ candidate in the
$\Upsilon(4S)$ rest frame. The events that satisfy the requirements, $\mb > 5.2$
GeV\cs{} and $|\de| <0.25$ GeV are selected for further analysis.

For events with multiple $B$ candidates, the best candidate is selected 
based on the $\chi^2$ of a vertex fit, that optimizes a vertex for all 
charged tracks 
in the final state, and a mass $\chi^2$. 
The latter term is necessary because of the photons in the \etap , 
which may introduce additional multiple candidates even for the same
set of charged particles. The
formula used to calculate the $\chi^2$ is:
\beq
\chi^2 = \chi^2_{\text{vertex}} + 
		\large[(M(\etap)-m_{\etap})/\sigma_{\etap} \large]^2 ,
\eeq
with $\chi^2_{\text{vertex}}$ being the $\chi^2$ from the charged 
particle vertex fit,
$M(\etap)$ the \etap{} candidate mass and $m_{\etap}$ the nominal mass of the
\etap{} and $\sigma_{\etap}=0.008$ GeV\cs{} 
the width of the \etap{} mass distribution.
About 10\% of events have multiple candidates and of these 10\%
are due to multiple photons.

Several event shape variables (defined in the center of mass frame) 
are used to distinguish
the more spherical \BB{} topology from 
the jet-like $q\bar q$ continuum events. 
The thrust angle $\theta_T$ is defined
as the angle between \etap{} momentum direction and 
the thrust axis formed by all tracks not from the same $B$
meson.
Jet-like events tend to peak near $|\cost| = 1$,
while spherical events have a flat distribution.
The requirement $|\cost| < 0.9$ is applied prior to any other 
event topology selections. 

Additional continuum suppression is obtained by using modified 
Fox-Wolfram moments~\cite{SFW} and the angle $\theta_B$ between the flight 
direction of the reconstructed $\Bbaro$ candidate and the beam axis. A 
Fisher discriminant (\FD)~\cite{fisher:1936} is formed by a linear combination
of $\cost$, $\sperp$ and five modified Fox-Wolfram moments. 
$\sperp$ is the ratio of
the scalar sum of the transverse momenta of all tracks outside a 
$45^{\circ}$ cone around the $\etap$ direction 
to the scalar sum of their total momenta.
Probability density functions (PDFs) are obtained from signal and 
background MC data samples.
These variables are then combined to form a event topology likelihood function
$\LK_c = $P$_c(\cosb) \cdot P_c(\FD)$,
where $P_c = $ PDF of signal ($s$) or continuum background $(q\bar{q})$.
Signal follows a $1-\cos^2\theta_B$ distribution while continuum background is
uniformly distributed in $\cosb$.
We select signal-like events by requiring a likelihood ratio
$\LR = \LK_{s}/(\LK_{s} + \LK_{q\bar q})$ criteria optimized by MC studies to 
suppress continuum background. For channels with an \erg{} decay an additional
variable $\cos\theta_{\hel}$, which is the angle between 
the $\etap$ momentum and the direction of one of
the decay pions in the $\rho$ rest frame, is
included for better signal-background separation. 

Further background discrimination is provided by the  quality of
the $B$ flavor tagging of the accompanying $B$ meson. We use the
standard Belle $B$ tagging package~\cite{TaggingNIM}, which gives the $B$ flavor
and a tagging quality $r$ ranging from zero for no flavor and unity
for unambiguous flavor assignment.  We divide the data into three
$r$ regions and separately optimize the \LR{} requirements using
signal and continuum background Monte Carlo samples.

The signal yields ($N_S$) are extracted using extended unbinned 
maximum-likelihood fits to two-dimensional (\de ,\mb ) distributions.
An extended likelihood function is:
\begin{eqnarray}
& & L(N_S,N_{B_j}) = \hspace*{5cm} \nonumber \\
& & \frac{e^{-(N_S+\sum_jN_{B_j})}}{N!} \prod_{i=1}^{N}
\left[N_{S} P_S(\de_i,M_{\mbox{\scriptsize bc}_i}) + 
	\sum_j N_{B_j} P_{B_j}(\de_i,M_{\mbox{\scriptsize bc}_i})\right]
\label{eq:ns-lkhd}
\end{eqnarray}
where $N$ is the total number of events, $i$ is an index running over the
events
and $P_S$ and $P_{B_j}$ are the PDFs for
signal and background, respectively, and the index $j$ runs over 
all background sources. 
The signal yield $N_S$ and background contents $N_{B_j}$ are determined
by maximizing the $L(N_S,N_{B_j})$ function, 
where the variable $N_{B_j}$ defines a $j$-dimensional submanifold of all
different backgrounds. 

To take into account the efficiency differences in two
data sets, 
$N_S$ is calculated as:
\beq 
\label{eq:NS}
N_S =  \eff_1 \, N_{\BB_1} \, \BF +  \eff_2 \, N_{\BB_2} \, \BF ,
\eeq
where $\BF$ is a branching fraction, and the $\epsilon_i$ and
$N_{\BB_i}$ are the efficiency and  the number of \BB{} pairs
for Set I and Set II.  We assume that the numbers of $B^+B^-$ and $B^0\Bbaro$
pairs are equal.

For charged $B$ decays, we divide the data into two samples for
positive and negative charges and extract the charge
asymmetry $A_{CP}$ in addition to the branching fraction \BF , using the
formula:
\beq
\label{eq:NK}
N_{\pm} = 0.5 \, ( 1 \mp A_{CP} ) \, N_S ,
\eeq
with $N_{+}$ ($N_{-}$) the number
of positively (negatively) charged kaons or pions and $N_S$ the number of 
signal events. 

The PDF shapes for each contribution are determined by MC studies.
We assume the signal shapes for \de{} and \mb{} to be independent and 
model the signal using a Gaussian with an
exponential tail (Crystal Ball line)~\cite{CBline} 
plus a Gaussian for \de{} and a Gaussian with an exponential
tail for \mb. 
The shape parameters are fixed from the signal MC sample, except for the means
and widths of both \de{} and \mb{} distributions, 
which are determined in the fit of
\btepk{} for decays with $K^+$ or $K_S^0$ in the final state for each data set 
(Set I and Set II) independently.
For \bteppip{} we use the parameters obtained from the fit to the charged kaon
mode. 

We consider up to four types of backgrounds separately in the fit:
continuum, $b\to c$, two types of charmless decays mentioned below. 
The \de{} and \mb{} distributions for continuum
backgrounds are found to be largely uncorrelated and are thus modeled with two 
independent one-dimensional functions. For \de{} we assign a first or second
order polynomial and the Argus function~\cite{bib:ARGUS} 
is used to model the \mb{} distribution. Charmless
$B$ decays and $b\to c$ backgrounds are modeled with 2-dimensional smoothed 
histograms. 
The contributions from charmless $B$ decays are modeled with two smoothed
histograms, one for one decay that makes a large contribution  to the
background, and one for all other charmless decays.
For \btepkp , the dominant decay decay mode that is modeled
separately is $B\to \etap K^*$;
for \btepks{} it is $B\to \rho^0 K_S^0$; and for \bteppip{} it
is the \btepkp{} feeddown.
The feeddown in \bteppip{} is modeled
with the same PDFs as used for the signal, shifted in \de{} and with an
additional correction factor for the change in the width in \de .

\begin{figure}[!htb]
\unitlength1.0cm
\centerline{
\epsfxsize 2.5 truein \epsfbox{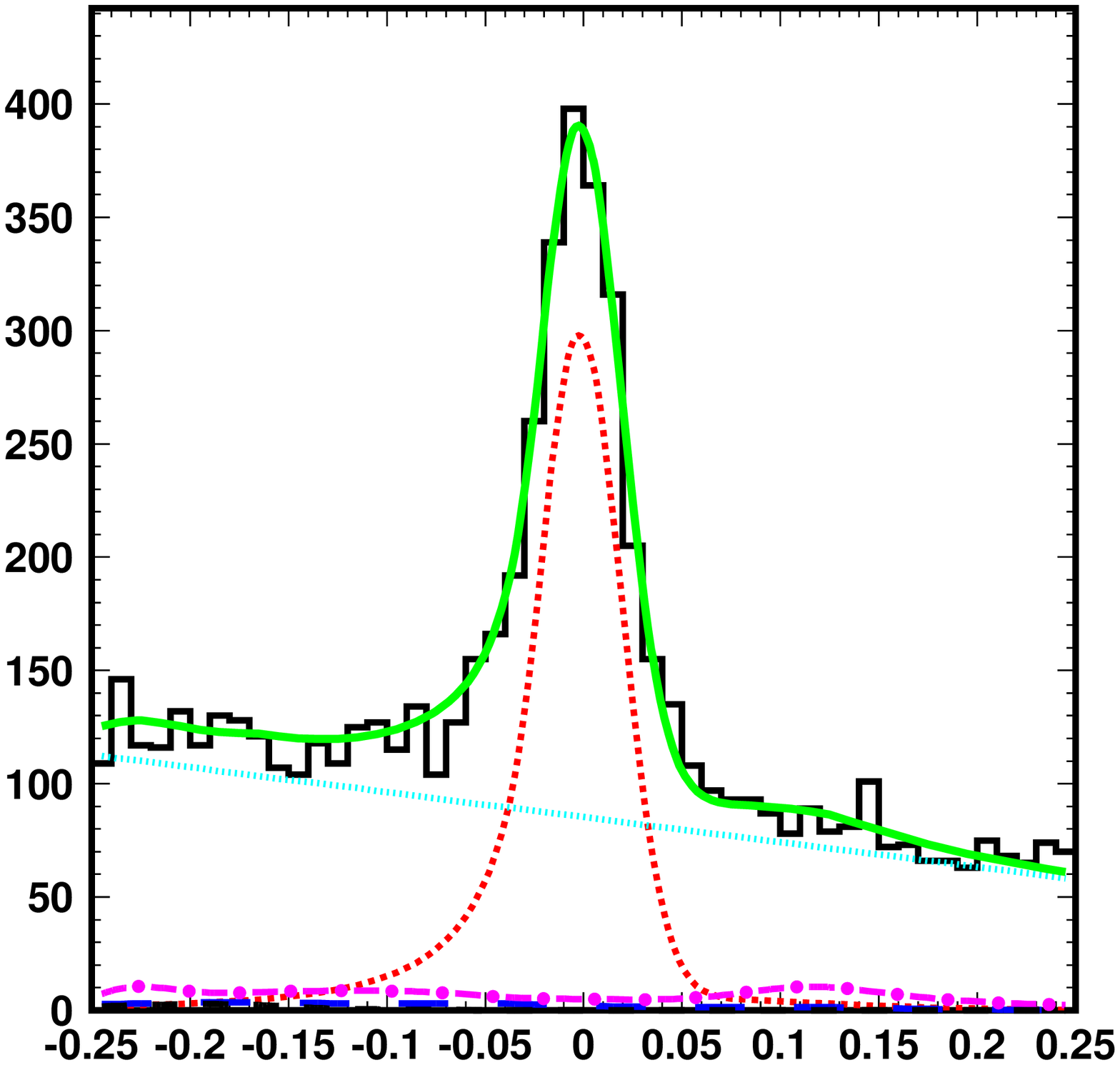}
\epsfxsize 2.5 truein \epsfbox{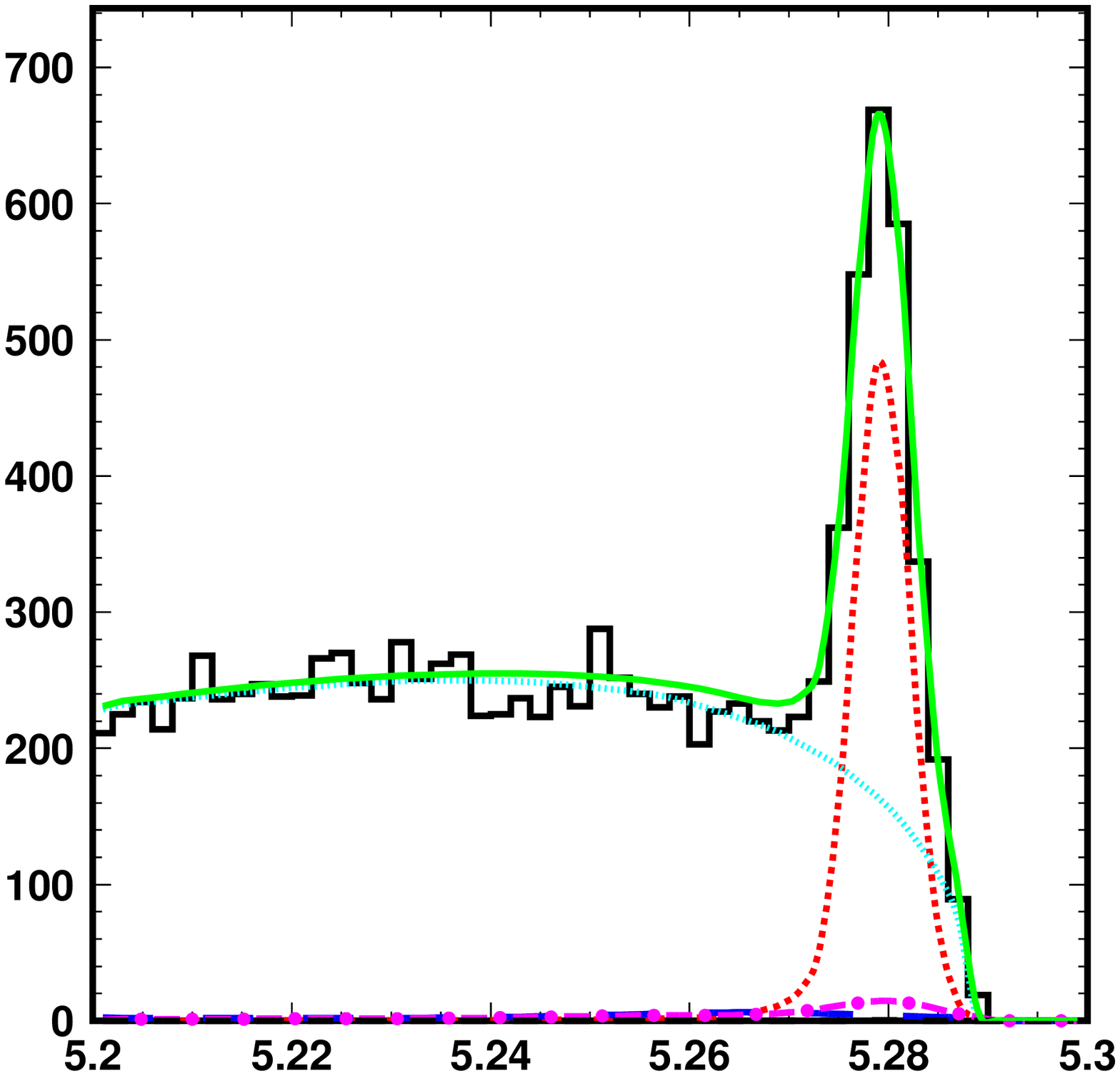}
\put(-10.3,0.0){\footnotesize{\sf\shortstack[c]{\de{} [GeV]}}}
\put(-3.5,0.0){\footnotesize{\sf\shortstack[c]{\mb{} [GeV]}}}
\put(-11.7,5.4){\footnotesize{\sf\shortstack[c]{a)}}}
\put(-5.2,5.4){\footnotesize{\sf\shortstack[c]{b)}}}
}
\caption{\label{fig:K} a) \de{} and b) \mb{} distributions for \btepkp{} 
(for \epp{} and \erg{} combined) for the
region $\mb>5.27$ GeV and $-0.1$ GeV $<\de<0.06$ GeV, respectively. 
The histograms represent data, 
the red small dashed line the signal contribution, 
the light blue dotted line continuum background, 
the dark blue large dashed line $b\to c$ backgrounds and 
the pink dash-dotted line charmless $B$ contributions. 
The green solid line is the sum of all contributions to the fit.}
\end{figure}

\begin{figure}[!htb]
\unitlength1.0cm
\centerline{
\epsfxsize 2.5 truein \epsfbox{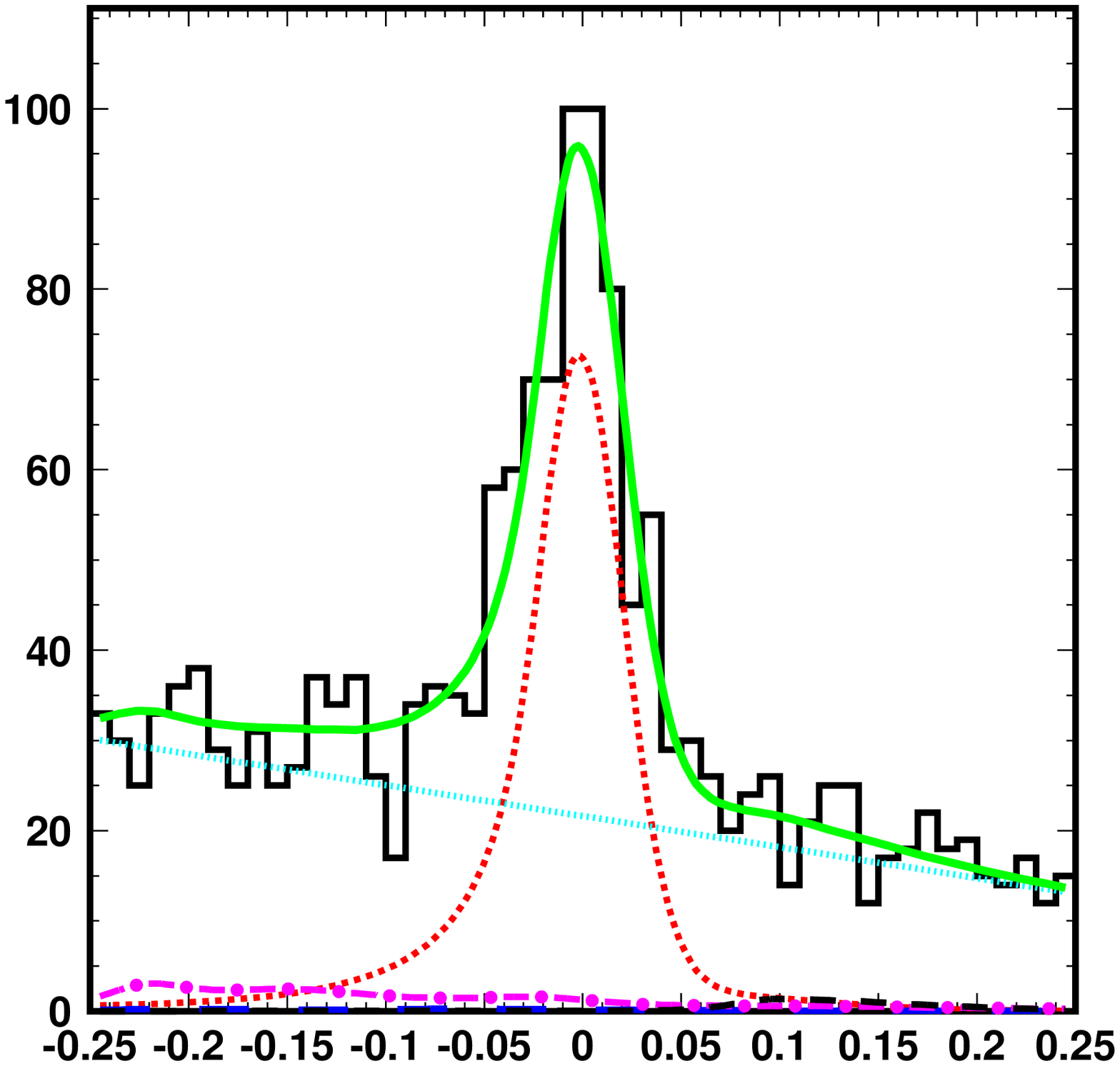}
\epsfxsize 2.5 truein \epsfbox{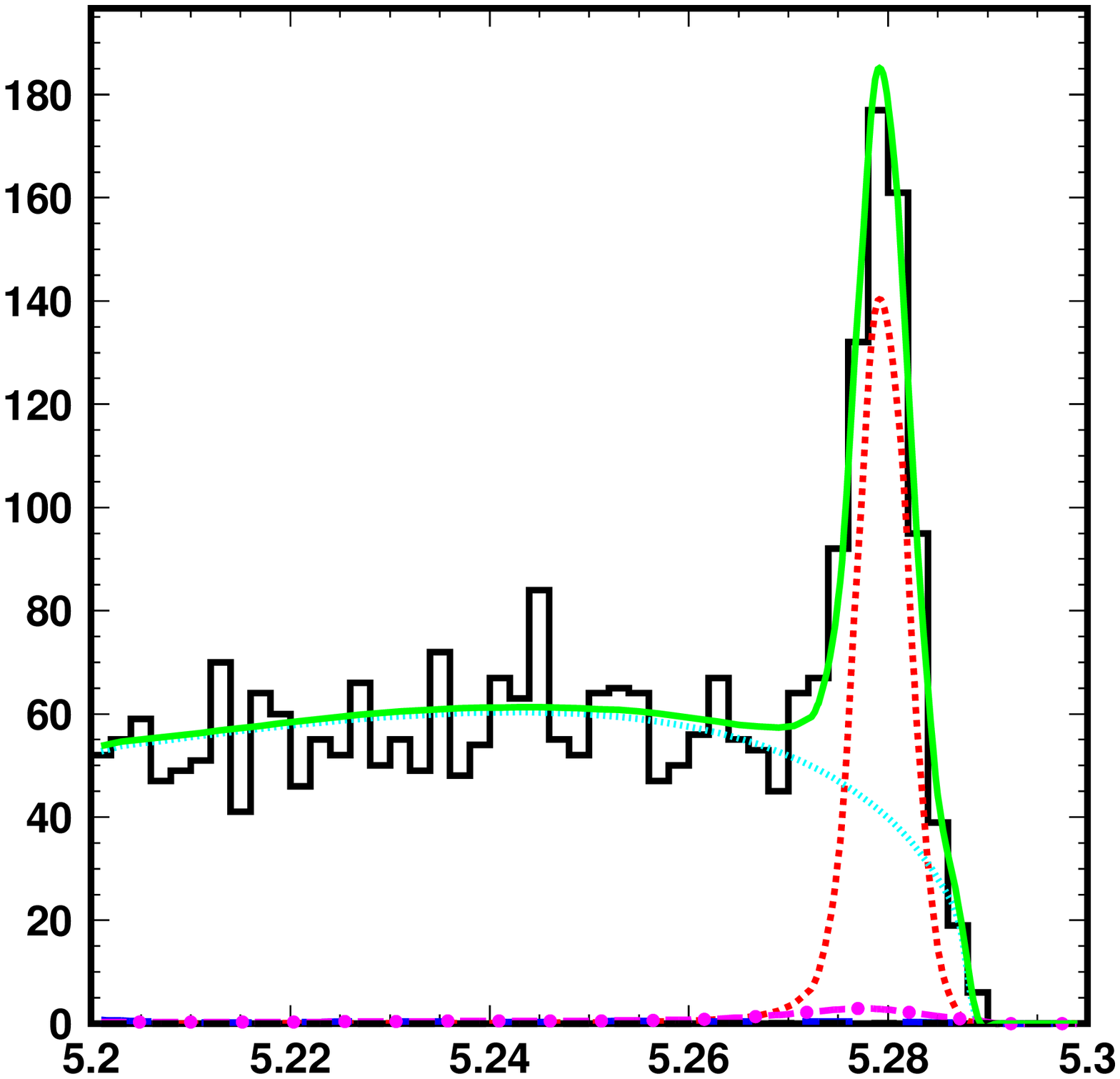}
\put(-10.3,0.0){\footnotesize{\sf\shortstack[c]{\de{} [GeV]}}}
\put(-3.5,0.0){\footnotesize{\sf\shortstack[c]{\mb{} [GeV]}}}
\put(-11.7,5.4){\footnotesize{\sf\shortstack[c]{a)}}}
\put(-5.2,5.4){\footnotesize{\sf\shortstack[c]{b)}}}
}
\caption{a) \de{} and b) \mb{} distributions for \btepks{}  
(for \epp{} and \erg{} combined) for the
region $\mb>5.27$ GeV and $-0.1$ GeV $<\de<0.06$ GeV, respectively. 
The histograms represent data, 
the red small dashed line the signal contribution, 
the light blue dotted line continuum background, 
the dark blue large dashed line $b\to c$ backgrounds and 
the pink dash-dotted line charmless $B$ contributions. 
The green solid line is the sum of all contributions to the fit.}
\label{fig:ks} 
\end{figure}

\begin{figure}[!htb]
\unitlength1.0cm
\centerline{
\epsfxsize 2.5 truein \epsfbox{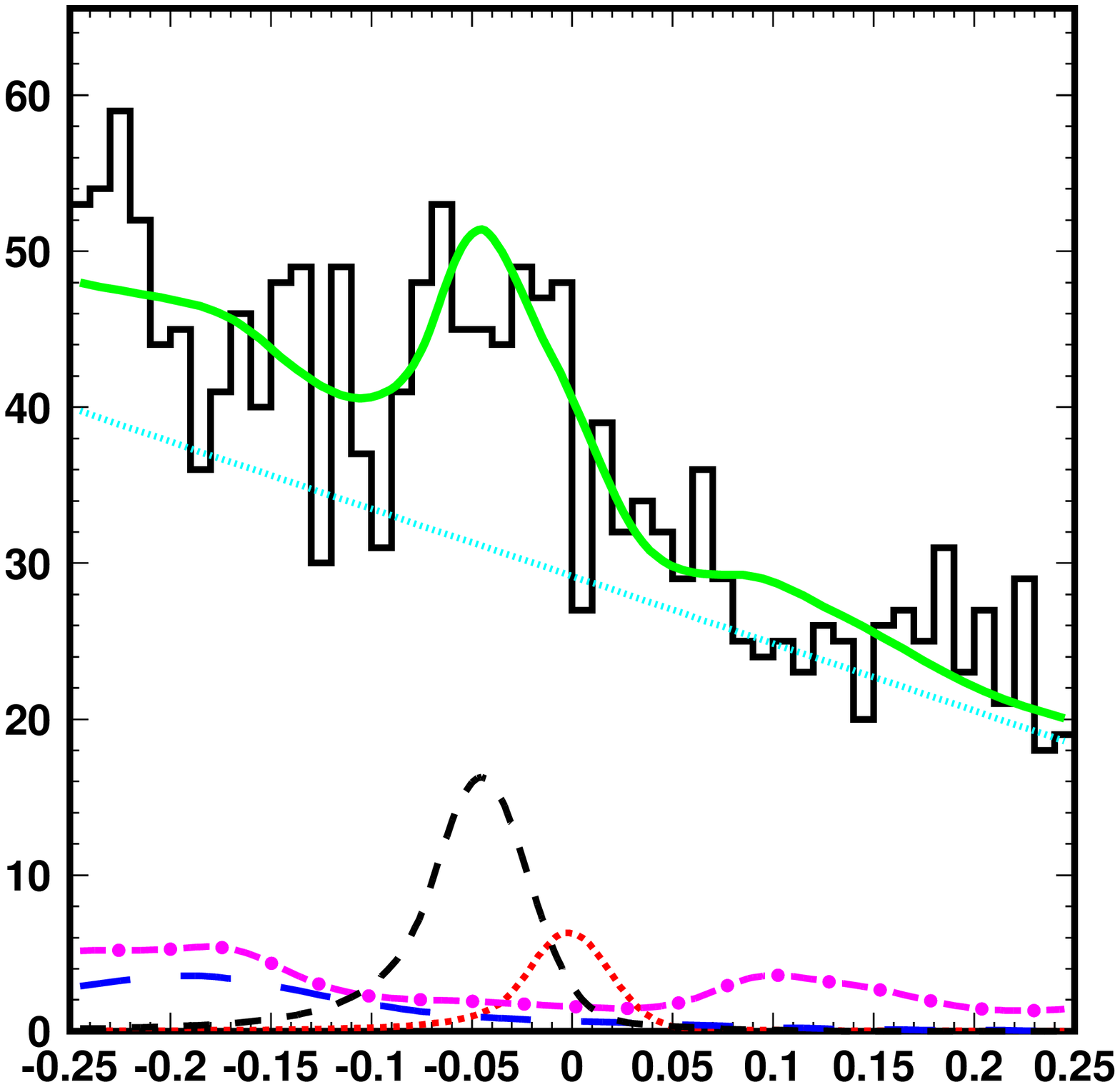}
\epsfxsize 2.5 truein \epsfbox{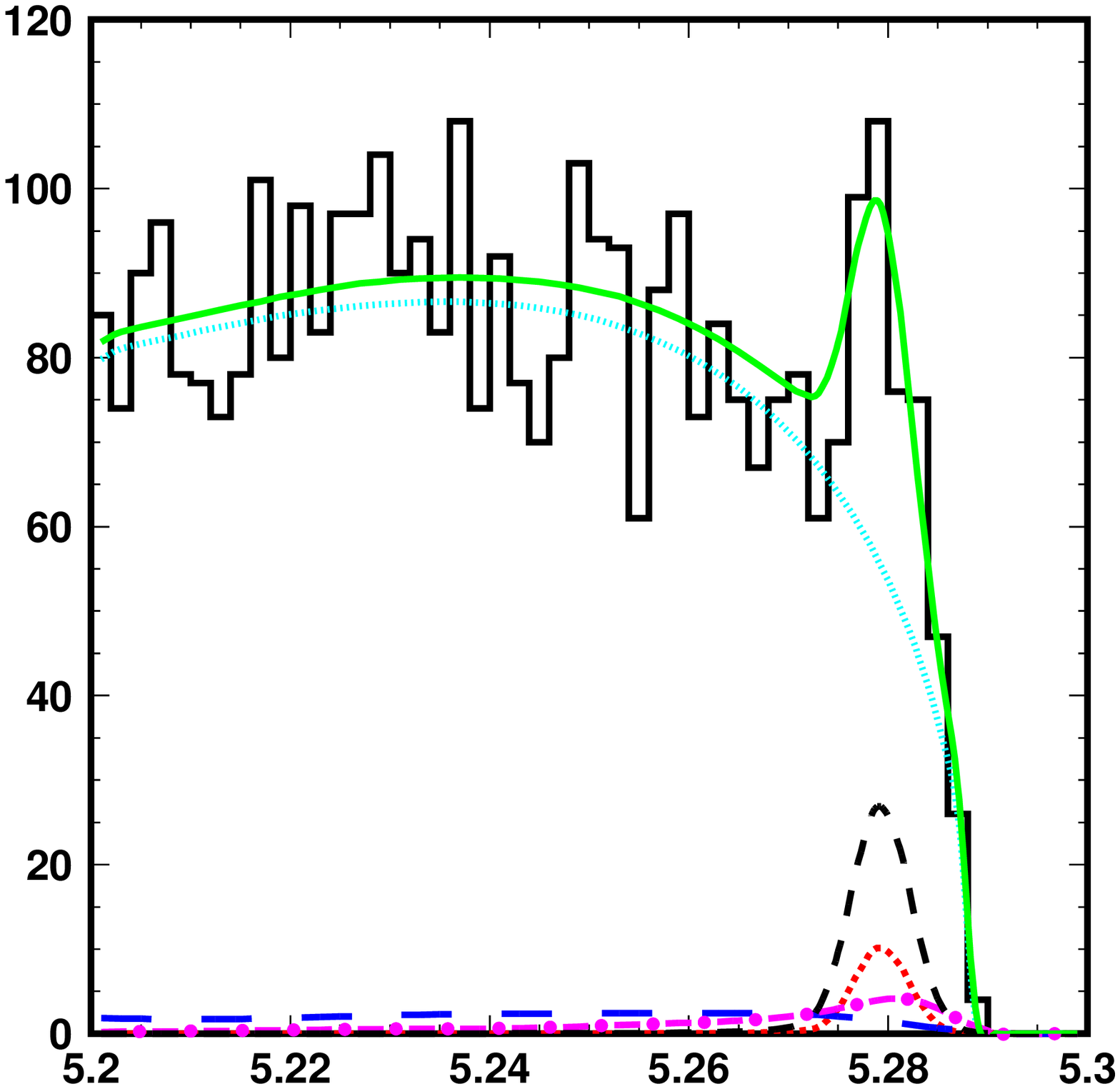}
\put(-10.3,0.0){\footnotesize{\sf\shortstack[c]{\de{} [GeV]}}}
\put(-3.5,0.0){\footnotesize{\sf\shortstack[c]{\mb{} [GeV]}}}
\put(-11.5,5.4){\footnotesize{\sf\shortstack[c]{a)}}}
\put(-5.2,5.4){\footnotesize{\sf\shortstack[c]{b)}}}
}
\caption{\label{fig:pi} a) \de{} and b) \mb{} distributions for \bteppip{}   
(for \epp{} and \erg{} combined) for the
region $\mb>5.27$ GeV and $-0.1$ GeV $<\de<0.06$ GeV, respectively. 
The histograms represent data, 
the red small dashed line the signal contribution, 
the light blue dotted line continuum background, 
the dark blue large dashed line $b\to c$ backgrounds, 
the pink dash-dotted line charmless $B$ contributions and 
the black medium dashed line \btepkp{} feeddown. 
The green solid line is the sum of all contributions to the fit.}
\end{figure}

The number of signal and background events are free parameters in the 
maximum likelihood fit.
The signal mean and width shape parameters for kaonic decays and the
continuum shape parameters are also free in all fits. 
For \epp{} modes our background MC studies show that no contributions from 
$b\to c$ decays are expected. For \btepk{} charmless backgrounds are 
also not expected.
For \bteppip{} there is a small contribution from charmless $B$ decays in
addition to the \btepkp{} feeddown.
For \erg{} modes, we float the number of events for $b\to c$
and two charmless $B$ contributions.
Other parameters are fixed to values determined from MC or
sideband data studies. 
For the combined results, we use a simultaneous fit with the branching fraction
and the charge asymmetry as common parameters for \epp{} and \erg. 
The projection plots of the fits are shown in
Fig.~\ref{fig:K}-\ref{fig:pi}. The reconstruction efficiencies for each decay 
and results of the fits are displayed in
Table~\ref{tab:ohwell}. The upper part of the table displays fit results for
individual fits for the two subdecay modes \epp{} and \erg. 
In addition, we list the charge asymmetry of the continuum background 
obtained from the fits. 
Since continuum background is thought to be distributed equally
for both charges, any significant observed asymmetry would indicate
a bias and is therefore included in the systematic error (see
below).

We find the significance of \bteppip{} yield is 3.0,
which is calculated as $\sigma = \sqrt{2\ln(L_{\rm max}/L_0)}$,
where $L_{\rm max}$ and $L_0$ denote
the maximum likelihood value and the
likelihood value at zero branching fraction, respectively.
The systematic error (mentioned below)
is included in the significance calculation 
by substracting the systematic error from the obtained branching 
fraction and recalculating the significance.

The reconstruction efficiencies are determined from signal MC samples before
including subdecay branching fractions and are
around 16-25\% for decays with \epp{} and 9-12\% for \erg{} before considering
subdecay branching fractions. 
The efficiencies are calculated separately for
both Set I and Set II. The efficiency for Set II is typically about 0.5\%
larger than for Set II. Correction factors due to differences
between data and MC are included for the charged track
identification, photon, $\pi^0$ and $\eta$ reconstruction, 
resulting in a correction factor of $\sim 0.9$. 
The corrections were determined from detailed studies that are
discussed in the section on the systematic error below.

We calculate a goodness of fit ($gof$)
based on binned projections of the 2-dim. fit
into \de.
We define:
\beq
gof = \chi^2 / dof, 
\eeq
with $\chi^2 = \Sigma^N_{i=1} \frac{(n_i -
\nu_i)^2}{\nu_i}$ the sum over all bins of the projections 
with $n_i$ and $\nu_i$ the number of
expected and observed events in each bin,
and $dof$ the degrees of
freedom.
Since we use an unbinned simultaneous fit in \de{} and \mb{} 
the projected $gof$ value is not a rigourous estimator but can
still be used to qualitatively evaluate how well we fit the data.
\begin{table}[htb]
\caption{Signal efficiencies with ($\epsilon_{\text{tot}}$) and without
($\epsilon$) subdecay
branching fractions included and averaged for Set I and Set II for \epp{}
and \erg , branching ratios \BF , asymmetry \ACP{} for signal and background
and goodness of fit ($gof$).
The errors in $\epsilon$ are statistical uncertainties in the MC while those
in $\epsilon_{\text{tot}}$ include errors of secondary branching
fractions.  For others, the first errors are statistical and
the second (if given) are systematic errors.}
\label{tab:ohwell}
\begin{tabular}
{@{\hspace{0.5cm}}l@{\hspace{0.5cm}}||@{\hspace{0.5cm}}c@{\hspace{0.5cm}}
@{\hspace{0.5cm}}c@{\hspace{0.5cm}}@{\hspace{0.5cm}}c@{\hspace{0.5cm}}}
\hline \hline
		&	\btepkp		& $B^0\to \etap K^0$	& \bteppip	\\
\hline
$\epsilon(\epp)$ [\%] 	& $25.23\pm 0.11$ & $19.9\pm 0.16$ & $16.3\pm 0.10$ \\
$\epsilon_{\text{tot}}(\epp)$ [\%] & $4.41\pm 0.14$ & $1.20\pm 0.04$ & $2.85\pm 0.10$ \\
yield (\epp)		&	$1140.7\pm 43.9$ & $243.2\pm 21.9$ & $17.4\pm 7.0$ \\
\BF(\epp) [$10^{-6}]$  	& $67.1{}\pm 2.6{}$ & 
				$52.5^{+4.8}_{-4.7}$ &
				$1.58^{+0.78}_{-0.70}$	\\
\ACP(\epp)	     	& $-0.003\pm 0.036$ &
				$ - $	& $0.25 ^{+0.49}_{-0.47}$ \\
\hline
$\epsilon(\erg)$ [\%] 	& $10.06 \pm 0.09$ & $11.6\pm 0.19$ &  $9.5\pm 0.08$ \\
$\epsilon_{\text{tot}}(\erg)$ [\%] & $2.97 \pm 0.10$ & $1.18\pm 0.04$ & $2.80\pm 01.0$ \\
yield (\erg)	&	$823.3\pm 42.9$	& $278.3\pm 24.1$ & $23.9\pm 10.1$ \\
\BF(\erg) [$10^{-6}]$   & $71.9{}^{+3.8}_{-3.7}$ & 
				$61.1^{+5.4}_{-5.2}$ &
				$2.21 \pm 1.28$	\\
\ACP(\erg)	     	& $0.09\pm 0.05$ &
				$ - $	& $-0.09 ^{+0.66}_{-0.74}$ \\
\hline\hline
yield		&	$1952.2\pm 60.6$ & $519.9\pm 32.3$ & $37.8\pm 14.5$ \\
\BF [$10^{-6}]$	      	& $68.6{}\pm 2.1{}\pm 3.6$ & 
				$56.6^{+3.6}_{-3.5}{}^{+3.3}_{-3.2}$ &
				$1.73^{+0.69}_{-0.63}\, \pm 0.12$	\\
\ACP		     	& $0.029\pm 0.028\pm 0.02$ &
			   $ - $ & $0.15^{+0.39}_{-0.38}{}^{+0.02}_{-0.06}$ \\
\ACP (continuum)	& $-0.013^{+0.008}_{-0.006}$ &  $-$ & 
						$0.018^{+0.006}_{-0.023}$ \\ 
$gof$			& 1.14 & 1.51 &	1.16	\\
\hline \hline
\end{tabular}
\end{table}
%
%
%

Systematic errors are estimated with various high statistics data samples. The
sources and their contributions are listed in Table~\ref{tab:syst}.
The dominant sources of background are the uncertainties 
of the reconstruction efficiency of charged tracks, the uncertainties in the
reconstruction efficiencies for $\eta$ mesons and photons and 
the uncertainty of the PDFs shape and other parameters, which
are estimated by varying each parameter of the PDFs by the $1 \sigma$
uncertainty
in its nominal value. The changes in the yield are added in quadrature.
The normalization of the $K$ feeddown to \bteppip{} is estimated by 
varying the assumed branching
fraction for the feeddown by the $\pm 1 \sigma$ uncertainty 
in the present measurement of
\btepkp{}, the error arising from differences in the fake efficiency is included
in the systematic error for the PDF parameters.
Corrections for \de{} and \mb{} differences between data and MC for the feeddown
shape of $\etap K$ to \bteppip{} are considered by varying the corrections by
one standard deviation and refitting.
Systematics arising from the \LR{} selection are studied by varying the
\LR{} selection and by a large $B\to D \pi$ sample, we conservatively use the
larger error.
The uncertainty of the subdecay branching fractions is given in the PDG.
The number of \BB{} mesons produced at Belle is estimated from
the number of continuum-subtracted hadronic events. The uncertainty
in this value is included in the systematic error.
The uncertainty of the particle identification is estimated with $D^{*+}\to D^0
\pi^+$ decays.
Other efficiency systematic errors are found to add up to less than 1\% and are
included with a 1\% error. 
All contributions are added in quadrature and 
we find the systematic errors for the three decays to lie between 5\% and 7\%.
For the charge asymmetry, efficiency based systematic errors cancel 
out and therefore we use the high statistics mode \btepkp{} to calculate
systematic errors and use the same absolute numbers for \bteppi.
The dominant contribution to the \ACP{} systematic error is from
the possible difference in the detector response for positive
and negative charged particles. We estimate this from the continuum
asymmetry to be 0.02. Other contributions from fitting and
normalization together result in a systematic error of 0.002.
Thus the total systematic error for the \ACP{} measurement is 0.02.
For \bteppip{} we assign an additional source of error by adding an asymmetry to
the \btepkp{} feeddown contribution. We find an error of $-0.06$, which is added
in quadrature to the other errors.
\begin{table}[hbtp]
\begin{center}
\caption{A breakdown of systematic uncertainties in percentage 
for branching fraction measurements. 
Systematic errors from
tracking, $\eff_{\gamma,\pi^0,\eta}$, the combined systematics from these three
particles, particle ID, $\eff$, the
systematics from efficiency calculations, \LR{} selection,
PDF, from varying the fit parameters by one $\sigma$,
\de / \mb{} corrections, which are varied by $\pm 1 \sigma$, 
the \btepkp{} feeddown to \bteppip , $\BF_s$, the subdecay branching
fraction errors and $N_{\BB }$, the uncertainty of the number of \BB , are
listed.}
\label{tab:syst}
\begin{tabular}{l||ccccccccccc}
\hline\hline
	&  \multicolumn{3}{c}{$\Delta \BF / \BF$ [\%]} \\
Source			& \btepkp	& \bteppip	& \btepks	\\
\hline
tracking 		& 3		& 3		& 4	\\
$\eff_{\gamma,\pi^0,\eta}$ & 3		& 3		& 3	\\
part. ID	 	& 0.7		& 0.9		& 0.6	\\
$\eff$ 			& 1		& 1		& 1	\\
\LR	 		& 2		& $^{+1.9}_{-2.7}$ 	& 2 \\
PDF			& $^{+0.5}_{-0.7}$ 	& $^{+2.6}_{-2.0}$ & 
							$^{+1.2}_{-1.3}$\\
\de / \mb  		& ---		& $^{+2.1}_{-1.9}$ & ---\\
$\etap K^+$ feeddown	& ---		& $^{+3.5}_{-2.9}$ & ---\\
$\BF_s$ 		& 1.5		& 1.5		& 1.5	\\
$N_{\BB}$ 		& 1		& 1		& 1	\\
\hline
Total			& 5.2		& $^{+7.1}_{-6.8}$ & 5.9 \\
\hline\hline
\end{tabular}
\end{center}
\end{table}

In summary, we report improved measurements with 35 times more
statistics of the charged and neutral \btepk{} decay. We find 
$3\sigma$ evidence for \bteppip{} and report the charge asymmetry for the 
decay modes \btepkp{} and \bteppip{}.
The central values of our branching fraction measurements are below
current PDG values, but are consistent within statistical errors.

We thank the KEKB group for the excellent operation of the
accelerator, the KEK cryogenics group for the efficient
operation of the solenoid, and the KEK computer group and
the National Institute of Informatics for valuable computing
and Super-SINET network support. We acknowledge support from
the Ministry of Education, Culture, Sports, Science, and
Technology of Japan and the Japan Society for the Promotion
of Science; the Australian Research Council and the
Australian Department of Education, Science and Training;
the National Science Foundation of China under contract
No.~10175071; the Department of Science and Technology of
India; the BK21 program of the Ministry of Education of
Korea and the CHEP SRC program of the Korea Science and
Engineering Foundation; the Polish State Committee for
Scientific Research under contract No.~2P03B 01324; the
Ministry of Science and Technology of the Russian
Federation; the Ministry of Higher Education, 
Science and Technology of the Republic of Slovenia;  
the Swiss National Science Foundation; the National Science Council and
the Ministry of Education of Taiwan; and the U.S.\
Department of Energy.


%


\begin{thebibliography}{99}

 
\bibitem{Grossman:1996ke}
Y.~Grossman and M.~P. Worah,
{\em Phys. Lett.} {\bf B395} (1997)
241,
hep-ph/9612269",

\bibitem{Atwood:1997bn}
D.~Atwood and A.~Soni, 
{\em Phys. Lett.} {\bf B405} (1997) 150,
hep-ph/9704357.
 
\bibitem{Kou:1999tt}
E.~Kou,
{\em Phys. Rev.}, {\bf D63} (2001) 054027,
hep-ph/9908214.

\bibitem{CC}
Throughout this paper, 
the inclusion of the charge conjugate mode decay is implied
unless otherwise stated.

\bibitem{Richichi:1999kj}
S.~J. Richichi {\em et al.} (CLEO Collaboration), 
{\em Phys. Rev.
  Lett.} {\bf 85} (2000) 520,
hep-ex/9912059.
 
\bibitem{Abe:2001pf}
K.~Abe {\em et al.} (Belle Collaboration), 
{\em Phys.
  Lett.} {\bf B517} (2001) 309,
hep-ex/0108010.
 

\bibitem{Aubert:2003bq}
B.~Aubert {\em et al.} (BABAR Collaboration),
  {\em Phys. Rev. Lett.} {\bf 91} (2003) 161801, hep-ex/0303046.

\bibitem{Ali:1998eb}
A.~Ali, G.~Kramer, and C.-D. Lu, 
{\em Phys. Rev.} {\bf D58}
  (1998) 094009,
hep-ph/9804363.
 
 
\bibitem{Chen:1999nx}
Y.-H. Chen, H.-Y. Cheng, B.~Tseng, and K.-C. Yang, 
{\em Phys. Rev.} {\bf D60} (1999) 094014,
hep-ph/9903453.
 
\bibitem{Kou:2001pm}
E.~Kou and A.~I. Sanda, 
{\em  Phys. Lett.} {\bf B525} (2002) 240,
hep-ph/0106159.
 
\bibitem{Chiang:2001ir}
C.-W. Chiang and J.~L. Rosner, 
{\em Phys. Rev.} {\bf D65} (2002) 074035,
hep-ph/0112285.
 
\bibitem{Hou:1998wy}
W.-S. Hou and B.~Tseng, 
  {\em Phys. Rev. Lett.} {\bf 80} (1998) 434,
hep-ph/9705304.
 
\bibitem{Yuan:1997ts}
F.~Yuan and K.-T. Chao, 
{\em Phys. Rev.} {\bf D56} (1997) 2495,
hep-ph/9706294.
 
\bibitem{Du:1998hs}
D.-s. Du, C.~S. Kim, and Y.-d. Yang, 
{\em Phys. Lett.} {\bf B426} (1998) 133,
hep-ph/9711428.
 
\bibitem{Ahmady:1998fa}
M.~R. Ahmady, E.~Kou, and A.~Sugamoto, 
{\em Phys. Rev.} {\bf D58} (1998) 014015,
hep-ph/9710509.
                                                                                
\bibitem{Xiao:2001uh}
Z.-j. Xiao, W.-j. Li, L.-b. Guo, and G.-r. Lu, 
{\em Mod. Phys. Lett.}
  {\bf A16} (2001) 441,
hep-ph/0103152.
 
\bibitem{Dutta:2002as}
B.~Dutta, C.~S. Kim, and S.~Oh, 
{\em Nucl. Phys. Proc. Suppl.} {\bf 111} (2002) 273,
hep-ph/0207171.
 
\bibitem{Khalil:2003bi}
S.~Khalil and E.~Kou, 
{\em Phys.
  Rev. Lett.} {\bf 91} (2003) 241602,
hep-ph/0303214.
                                                                                
\bibitem{Dariescu:2004bs}
M.-A. Dariescu and C.~Dariescu, 
{\em Eur. Phys. J.} {\bf C36} (2004) 215,
hep-ph/0404148.
 
\bibitem{Kramer:1993yu}
G.~Kramer, and W.~F.~Palmer, and H.~Simma,
{\em Nucl. Phys.} {\bf B428} (1994), 77,
hep-ph/9402227,

\bibitem{KEKB}
S.~Kurokawa and E.~Kikutani, 
{\em Nucl. Instr. and. Meth.} {\bf A499} (2003) 1,
and other papers included in this volume.

\bibitem{Belle}
A.~Abashian {\it et al.} (Belle Collab.),
{\em Nucl. Instr. and Meth.} {\bf A479}, 117 (2002).

\bibitem{Ushiroda} Y. Ushiroda (Belle SVD2 Group),
{\em Nucl. Instr. and Meth.} {\bf A511} 6 (2003). 


 
\bibitem{bib:PDG04}
S.~Eidelman {\em et al.} (Particle Data Group), 
{\em Phys. Lett.} {\bf B592} (2004) 1.

\bibitem{SFW}
 The Fox-Wolfram moments were introduced in
 G.~C.~Fox and S.~Wolfram, {\em Phys. Rev. Lett.} {\bf 41}, 1581 (1978).
 The Fisher discriminant used by Belle, based on modified Fox-Wolfram
 moments (SFW), is described in 
 K.~Abe {\it et al.} (Belle Collab.), {\em Phys. Rev. Lett.} {\bf 87},
 101801 (2001) and
 K.~Abe {\it et al.} (Belle Collab.), {\em Phys. Lett.} {\bf B 511}, 151
 (2001). 

\bibitem{fisher:1936}
R.~A. Fisher, {\em Annals of Eugenics} {\bf 7} (1936) 179.
                                                                                
\bibitem{TaggingNIM}
H. Kakuno {\it et al.}, {\em Nucl. Instr. and Meth.} {\bf A533} 516 (2004). 

\bibitem{CBline}
J.E.Gaiser {\em et al.} (Crystal Ball Collaboration) {\em Phys. Rev.} {\bf D34}, 
711 (1986).

\bibitem{bib:ARGUS}
H.~Albrecht {\em et al.} (ARGUS Collaboration),
 {\em Phys. Lett.} {\bf B241} (1990) 278.
 


\end{thebibliography}
\end{document}